\documentclass[twoside]{article}
\usepackage{np}

\begin{document}

\runninghead {Gennadii A. Kotel'nikov}
{On the electrodynamics with faster-than-light motions}

\thispagestyle{empty}\setcounter{page}{71}
\vspace*{0.88truein}
\fpage{71}

\centerline{\bf ON THE ELECTRODYNAMICS WITH FASTER-THAN-LIGHT}

\vspace*{0.035truein}
\centerline{\bf MOTION}
\vspace*{0.035truein}

\vspace*{0.37truein}
\centerline{\footnotesize Gennadii A. Kotel'nikov}

\centerline{\footnotesize \it
RRC Kurchatov  Institute,  Moscow, 123182,  Russia}


\baselineskip 5mm

\vspace*{0.21truein}

\abstracts{The version of  electrodynamics  is  constructed  in  which
faster-than-light motions of fields and particles with real masses are
possible.}{}{}


\bigskip

$$$$

\section{Introduction}\label{1}
For certainty,  by faster-than-light motions we  will  understand  the
motions  with  velocities  $v>3\cdot10^{10}$ cm/sec.  The existence of
such motions is the question discussed in modern physics.  In spite of
the well-known  conservatism,  in  the  world   science   considerable
attention is given to this question.

Blokhintsev (1946-1947) [1]  paid  attention  to  the  possibility  of
formulating   the   field  theory  that  permits  the  propagation  of
faster-than-light (superluminal) interactions outside the light  cone.
Later  (1952)  [2]  he  also noted the possibility of the existence of
superluminal solutions in the nonlinear equations of electrodynamics.

\begin{sloppypar}
Kirzhnits (1954) [3] showed that a particle possessing the  tensor  of
mass      ${M^i}_k=diag(m_0,m_1,m_1,m_1),     \     i,k=0,1,2,3,     \
g_{ab}=diag(+,-,-,-)$ can move  with  the  faster-than-light  velocity
$v=cm_0/m_1>c$ if $m_0>m_1$.
\end{sloppypar}

Terletsky (1960) [4] introduced into theoretical physics the particles
with imaginary rest masses moving faster than light.

Feinberg (1967) [5] named these particles tachyons and described their
main properties.

Research on the superluminal  tachyon  motions  opened  up  additional
opportunities  which  were  studied  by  many authors,  for example by
Bilaniuk and Sudarshan~[6],  Recami [7],  Mignani [8],  Kirzhnits  and
Sazonov [9],  Corben [10],  Patty [11], Recami, Fontana and Caravaglia
[12]. It  has  led  to   original   scientific   direction   (hundreds
publications).  The  tachyon movements may formally be described by SR
expanded to the area of motions where  $s^2<0$.  For  comparison,  the
standard  theory  describes  motions on the zero cone $s^2=0$ at $v=c$
and in the area $s^2>0$ when $v<c$.

The publications are also known in which the violation  of  invariance
of  the  speed  of light is considered [13] - [19].  We can note,  for
example,  the Pauli's monograph [13] where the elements  of  Ritz  and
Abraham   theories   are   contained;   the  Logunov's~  "Lections  on
Fundamentals of Relativity Theory"  [14];  the  Glashow's  paper  [15]
discussing   the   experimental   consequences  of  violation  of  the
Lorentz-invariance in astrophysics;  publications [16] - [19]  on  the
violation of invariance of the velocity of light in SR.

Below a  version of the theory permitting faster-than-light motions of
electromagnetic fields and  charged  particles  with  real  masses  is
proposed as the continuation of such investigations.
\bigskip

\begin {sloppypar}\noindent\section{Space-Time Metric.  Transformation
Law of Coordinates}\end{sloppypar}
Let us start from the condition of  invariance  of  the  infinitesimal
space-time  interval of 4-space $R^4$,  the metric properties of which
may depend on the velocity of a particle being investigated. Supposing
that space  is  homogeneous  and  isotropic  we take the metric in the
form:
\begin{equation}\label{g49}
\begin{array}{c}
ds^2=({c_0}^2+v^2)dt^2-dx^2-dy^2-dz^2=                              \\
({c'_0}^2+v'^2)(dt')^2-(dx')^2-(dy')^2-(dz')^2 - invariant.
\end{array}
\end{equation}
Here $x,  y,  z$ are the spatial coordinates, \ $t$ is the time, $c_0,
c'_0$ are the proper values of the speed of light, $v$ is the velocity
of a particle with respect to the reference frame $K$.  Let us connect
the co-moving frame $K'$ with this particle.  Let the proper  speed  of
light be invariant
\begin{equation}\label{c0}
c_0=c'_0=3\cdot10^{10} cm/sec - invariant.
\end{equation}
In this  case,  as  follows  from (\ref{g49}),  the common time $t'_0$
similar to the Newton time may be introduced on the trajectory of  the
frame $K'$ movement with the velocity ${\bf v}=d{\bf x}/dt$:
\begin{equation}\label{lt}
dt=dt'_0 \to t=t'_0.
\end{equation}
The value  \fnm{a}\fnt{a}{In  the  form of $c'=c(1-\beta^2)^{1/2}$ the
expression (\ref{c}) was obtained by Abraham in  the  model  of  ether
[13].}
\begin{equation}\label{c}
\displaystyle
c=\pm c_0\sqrt{1+\frac{v^2}{{c_0}^2}}
\end{equation}
is the velocity of light corresponding to the particle  velocity  $v$.
Hereafter  let  us  agree to name the value $3\cdot10^{10}$ cm/sec the
speed of light,  the value  $c\ne  c_0$  the  velocity  of  light.  In
accordance  with  the  hypothesis  of  homogeneity and isotropy of the
space-time, the velocity $v$ of a free particle does not depend on $t$
and ${\bf x}$.  The velocity of light (\ref{c}) is a constant value in
this case. By substituting
\begin{equation}\label{nt}
x^0=\int\limits_{0}^{t}cd\tau=\pm\int\limits_{0}^{t}
c_0\sqrt{1+\frac{v^2}{{c_0}^2}} d\tau,
\end{equation}
we may   introduce   the  "time"  $x^0$  and  rewrite  the  expression
(\ref{g49}) in the form
\begin{equation}\label{nlt}
ds^2=(dx^0)^2-(dx^1)^2-(dx^2)^2-(dx^3)^2,
\end{equation}
\begin{sloppypar}\noindent
where $x^{1,2,3}=(x,y,z)$.   As   is  known,  the  metric  (\ref{nlt})
describes  the  flat  homogeneous  Minkowski  space-time  $M^4$   with
$g_{ik}=diag(+1,-1,-1,-1)$, \ $i,k=0,1,2,3$. The infinitesimal space -
time  transformations,  retaining   the   invariance   of   the   form
(\ref{nlt}) take the form [20]
\end{sloppypar}
\begin{equation}\label{f2}   
dx'^i={L^i}_k  dx^k,  \  i,k=0,1,2,3,
\end{equation} 
where ${L^i}_k$ is the matrix of  the  six-dimensional  Lorentz  group
$L_6$.   Let   us,   for  example,  write  the  matrix  ${L^i}_k$  for
one-dimensional  Lorentz  group  $L_1$  with   the   group   parameter
$\beta=V/c=const$ [20]:
\begin{equation}\label{2.6a}    
{L^i}_k=\left(     
\begin{array}{cccc}
\displaystyle        
\frac{1}{\sqrt{1-\beta^2}}&        
\displaystyle
-\frac{\beta}{\sqrt{1-\beta^2}}&0&0                                 \\
\displaystyle
-\frac{\beta}{\sqrt{1-\beta^2}}&                        
\displaystyle
\frac{1}{\sqrt{1-\beta^2}}&0&0                                     \\
\displaystyle    
0&0&1&0                                                            \\
\displaystyle 
0&0&0&1 
\end{array} \right).  
\end{equation} 
\begin{sloppypar}\noindent
Analogously to  (\ref{f2}),  we  may   introduce   the   formulae   of
transformation              of              the             4-velocity
$u^i=(dx^0/ds,dx^{\alpha}/ds)=(1/\sqrt{1-u^2},u^{\alpha}/\sqrt{1-u^2})
=(c/c_0,cu^{\alpha}/c_0)$, where    $ds=\sqrt{1-u^2}dx^0=(c_0/c)dx^0$,
$u^{\alpha}=v_{x,y,z}/c$,    $\alpha=1,2,3$,     $u^2=g_{ik}u^iu^k=1$:
\end{sloppypar}
\begin{equation}\label{U}    
u'^i={L^i}_ku^k, i,k=0,1,2,3.
\end{equation}  
As a  result,  the   one-dimensional   infinitesimal   transformations
corresponding to the given matrix, are:
\begin{equation}\label{f3}  
\begin{array}{c}  
\displaystyle
dx'^0=\frac{dx^0-\beta dx^1}{\sqrt{1-\beta^2}};                      \
dx'^1=\frac{dx^1-\beta  dx^0}{\sqrt{1-\beta^2}};                     \
dx'^2=dx^2;   \    dx'^3=dx^3;                                       \
\displaystyle
c'=c\frac{1-\beta u^1}{\sqrt{1-\beta^2}}.
\end{array} 
\end{equation} 
\begin{sloppypar}\noindent
Here the  latest  formula  follows  from   the   law   of   4-velocity
transformation   $u'^0={L^0}_ku^k$  under the   condition  (\ref{c0});
$c'=c_0/\sqrt{1-u'^2},  \ c=c_0/\sqrt{1-u^2}$,  \ $c'=\gamma  c$.  The
reciprocal transformations may be obtained by prime  permutation.  The
group  parameters  are  related  by  the  ratios  $\beta'=-\beta$  and
$\gamma'=\gamma^{-1}$. The   integral   homogeneous    transformations
corresponding to (\ref{f3}) are
\begin{equation}\label{if3}
\begin{array}{c}   
\displaystyle
x'^0=\frac{x^0-\beta x^1}{\sqrt{1-\beta^2}};                         \
x'^1=\frac{x^1-\beta x^0}{\sqrt{1-\beta^2}};   \   x'^2=x^2;         \
x'^3=x^3;                                                            \
c'=c\frac{1-\beta u^1}{\sqrt{1-\beta^2}}.
\end{array} 
\end{equation} 
They are  induced  by  the   operator   $X=x_1\partial_0-x_0\partial_1
-u^1c\partial_c$  which  is  the  sum of Lorentz group $L_1$ generator
$J_{01}=x_1\partial_0-x_0\partial_1$ and the generator $D=c\partial_c$
of scale transformations group $\triangle_1$.  These generators act in
5-space $M^4XV^1$ where $V^1$ is  a  subspace  of  the  velocities  of
light.  We  may say that the transformations (\ref{if3}) belong to the
group of direct product $L_1X\triangle_1$. The generators $J_{01}$ and
$D$  and  transformations  (\ref{if3})  are  respectively the symmetry
operators and symmetry transformations for the equation of  the  light
cone  $s^2$  in  the 5-space $V^5=M^4XV^1$ where $|c|<\infty$ includes
the subset $c_0<|c|<\infty$. We have
\begin{equation}\label{cone}
\begin{array}{c}
\displaystyle
s^2=(x^0)^2-(x^1)^2-(x^2)^2-(x^3)^2=0,                              \\
\displaystyle
J_{01}(s^2)=0, \ D(s^2)=0,                                          \
\displaystyle
[J_{01},D]=0.
\end{array}
\end{equation}
Below we shall consider the case of positive values of  the  speed  of
light. In this case SR is realized on the hyperplane $c=c_0$.
\end{sloppypar}
The relationships between the variables for the space-time $R^4$  with
metric   (\ref{g49})   and  for  Minkowski  space  $M^4$  with  metric
(\ref{nlt}) are as follows:
\begin{equation}\label{fg1}
\begin{array}{c}  
\vspace{2mm}
\displaystyle      
\frac{\partial}{\partial t}=\frac{\partial
x^0}{\partial t}\frac{\partial}{\partial x^0}+
\sum_{\alpha}\frac{\partial x^{\alpha}}{\partial t}\frac{\partial}
{\partial x^{\alpha}}=c\frac{\partial}{\partial x^0};               \\
\vspace{2mm}
\displaystyle \frac{\partial}{\partial x}=\frac{\partial x^0}{\partial
x}      \frac{\partial}{\partial      x^0}+\sum_{\alpha}\frac{\partial
x^{\alpha}}    {\partial    x}\frac{\partial}{\partial    x^{\alpha}}=
\Big(\int\frac{\partial c}{\partial x}d\tau\Big)
\frac{\partial }{\partial x^0}+
\frac{\partial}{\partial   x^1};                                    \\
\vspace{2mm}
\displaystyle
\frac{\partial}{\partial     y}=\frac{\partial     x^0}{\partial    y}
\frac{\partial}{\partial x^0}+\sum_{\alpha}\frac{\partial  x^{\alpha}}
{\partial y}\frac{\partial}{\partial x^{\alpha}}=
\Big(\int\frac{\partial c}{\partial y}d\tau\Big)
\frac{\partial }{\partial x^0}+
\frac{\partial}{\partial   x^2};                                    \\
\displaystyle
\frac{\partial}{\partial     z}=\frac{\partial     x^0}{\partial    z}
\frac{\partial}{\partial x^0}+\sum_{\alpha}\frac{\partial  x^{\alpha}}
{\partial z}\frac{\partial}{\partial x^{\alpha}}=
\Big(\int\frac{\partial c}{\partial z}d\tau\Big)
\frac{\partial }{\partial x^0}+
\frac{\partial}{\partial   x^3}.                                    
\end{array} 
\end{equation}
We restrict ourselves by studying a variant of the theory in which the
velocity  of light in the range of interactions may only depend on the
time $t$,   i.e.   $c=c(t)\leftrightarrow   c=c(x^0)$,    where    the
relationship between $x^0$ and $t$ may be deduced from the solution of
Eq. (\ref{nt}). Then
\begin{equation}\label{nabla}
\displaystyle
\nabla c(x^0)=0, \ c=c(x^0), \ u=u(x^0);      \to
\nabla c(t)=0, \ c=c(t), \ v=v(t).
\end{equation}

Let us note some features of motions in the space $M^4(x^0,{\bf x})$.

1.~As in SR,  the parameter $\beta=V/c$ in the present work is in  the
range $0\leq\beta<1$.

2.~As in SR, the value $dx^0$ is the exact differential in view of the
condition $\nabla c=0$.

3.~As distinct from SR,  the "time" $x^0=ct$ in the present work is  a
function of the time $t$ only for the case of a free particle.  In the
range of interaction the velocity of light depends on  time  $t$.  The
value  $x^0$  becomes the functional (\ref{nt}) of the function $c(t)$
and takes into consideration the history of moving the particle.

4.~The group parameter $\beta=V/c$ of the matrix (\ref{2.6a})  may  be
constant  not  only  at  the  constant velocity of light,  but also at
$c=c(t)$.        Indeed        we        may        accept        that
$0\leq\beta=V(t)/c_0(1+v^2(t)/{c_0}^2)^{1/2}=constant\leq 1$, which is
not in contradiction with $V=V(t),  \ c=c(t)$. Supposing here $V=0$ we
find $\beta=0$.  Also,  with $V\to v$, $v\to\infty$, we find $\beta\to
1$.  This property permits the using of the  matrix  (\ref{2.6a})  for
constructing  Lorentz  invariants  in  the  range of interaction where
$c=c(t)$.

Keeping this in mind,  let us construct in the Minkowski space $M^4$ a
theory like SR, reflect it on the space $R^4$ by means of the formulas
(\ref{fg1}) with $\nabla c=0$ and consider the main properties of  the
constructed theory.

\section{Action, Energy, Momentum}\label{4} 
Following [20],  analogously to SR we may construct  the  integral  of
action in the form:
\begin{equation}\label{2.26b}
\begin{array}{c}
\vspace{1mm}
\displaystyle
S={S}_m+{S}_{mf}+{S}_f=
\displaystyle
-m_0c_0\int ds - \frac{e}{c_0}\int A_i dx^i-
\frac{1}{16\pi c_0}\int F_{ik}F^{ik} d^4 x=                         \\
\vspace {1mm}
\displaystyle
\int[-m_0c_0\sqrt{1-u^2}+\frac{e}{c_0}({\bf A}\cdot{\bf u}-\phi)]dx^0-
\frac{1}{8\pi c_0}\int(E^2-H^2)d^3xdx^0=                           \\
\displaystyle
-m_0c_0\int ds - \frac{1}{c_0}\int A_ij^id^4x-
\frac{1}{16\pi c_0}\int F_{ik}F^{ik} d^4 x.
\end{array}
\end{equation}
\noindent
\begin{sloppypar}\noindent
Here in       accordance       with       [20]      ${S}_m=-m_0c_0\int
ds=-m_0c_0\int(c_0/c)dx^0=-m_0c_0\int(1-u^2)dx^0$ is the action for  a
free particle; ${S}_f=-1/16\pi c_0\int F_{ik}F^{ik}d^4x$ is the action
for     free     electromagnetic     field,     ${S}_{mf}=-(e/c_0)\int
A_idx^i=-(1/c_0)\int  A_ij^id^4x$  is  the action corresponding to the
interaction between the charge $e$ of a particle  and  electromagnetic
field;   ${\bf   u}={\bf  v}/c$  is  the  "velocity"  of  a  particle;
$A^i=(\phi,{\bf  A})$   is   the   4-potential;   \   $A_i=g_{ik}A^k$;
$j^i=(\rho,\rho{\bf u})$  is  the  4-vector  of  the  current density;
$\rho$  is  the   charge   density;   $F_{ik}=(\partial   A_k/\partial
x^i-\partial  A_i/\partial  x^k)$  is  the  tensor  of electromagnetic
field;  $i,k=0,1,2,3$; $g_{ik}=diag(+,-,-,-)$; ${\bf E}=-\partial {\bf
A}/\partial   x^0-\nabla\phi$   is   the   electrical   field;   ${\bf
H}=\nabla{\rm    X}{\bf     A}$     is     the     magnetic     field;
$F_{ik}F^{ik}=2(H^2-E^2)$;  $dx^4=dx^0dx^1dx^2dx^3$  is the element of
the invariant 4-volume.  The proper value of the speed of light $c_0$,
the  rest  mass  $m_0$,  the  electric  charge  $e$  are the invariant
constants of the theory. The integral of action (\ref{2.26b}) we shall
name the modified action.

In spite of the similarity,  the action (\ref{2.26b}) differs from the
action of  SR  [20].  The  current  density  has  taken  in  the  form
$j^i=(\rho,\rho{\bf    u})=(\rho,\rho    {\bf    v}/c)$   instead   of
$j^i=(\rho,\rho{\bf v})$ from  [20].  The  electromagnetic  field  has
taken  in  the  following  form  ${\bf  E}=-\partial  {\bf A}/\partial
x^0-\nabla\phi=-(1/c)\partial {\bf A}/\partial  t-\nabla\phi$  instead
of   the   expression   ${\bf   E}=-(1/c_0)\partial  {\bf  A}/\partial
t-\nabla\phi$ [20].  The current density (\ref{2.26b}) is  similar  to
the  current  density  from  Pauli's  monograph  [13]  with  the  only
difference that the 3-current density in (\ref{2.26b})  has  taken  in
the   form   $\rho{\bf   v}/c$  instead  of  $\rho{\bf  v}/c_0$  [13].
\end{sloppypar}

In addition to the Lorentz-invariance [20],  the action  (\ref{2.26b})
is  also invariant with respect to any transformations of the velocity
of light  and,  consequently,  with  respect  to  the  transformations
$c'=\gamma  c$,  as  the  value $c$ is not contained in the expression
(\ref{2.26b})  \fnm{b}\fnt{b}{The  choice  of  the   action   integral
(\ref{2.26b}) is ambiguous. Instead of (\ref{2.26b}), we may introduce
the action in the form  $cS$,  where  $S$  is  the  action  [20].  The
"momentum" $c{\bf  p}=m_0c_0{\bf v}$ and the energy ${\cal E}=m_0c_0c$
are the integrals of motion in this case.  The mass  of  movement  $M$
depends    on    the    velocity    $v$   accordingly   to   the   law
$M=m_0/(1+v^2/{c_0}^2)^{1/2}$  [17].}.  As   a   result   the   action
(\ref{2.26b})   is  invariant  with  respect  to  the  transformations
(\ref{if3}) from the group of direct  product  $L_1X\triangle_1\subset
L_6X\triangle_1$,  containing  Lorentz group $L_6$ and the scale group
$\triangle_1$ as subgroups.

\begin{sloppypar} The  modified  Lagrangian  $L$,  generalized   4   -
momentum ${\bf P}$ and generalized energy $H$ of a particle correspond
to the modified action (\ref{2.26b}). We have:
\end{sloppypar}
\begin{equation}\label{2.27}
L=-m_0c_0\sqrt{1-u^2}+\frac{e}{c_0}({\bf A}\cdot{\bf u}-\phi);
\end{equation}
\begin{equation}\label{2.27a}
{\bf P}=\frac{\partial  L}{\partial{\bf u}}=\frac{m_0c_0{\bf u}}{\sqrt
{1-u^2}}+\frac{e}{c_0}{\bf A}={\bf p}+\frac{e}{c_0}{\bf A}=m_0{\bf v}+
\frac{e}{c_0}{\bf A};
\end{equation}
\begin{equation}\label{2.28b}
H={\bf P}\cdot{\bf u}-L=\frac{m_0c_0c+
e\phi}{c_0}=\frac{{\cal E}+e\phi}{c_0}.
\end {equation}
Here
\begin{equation}\label{mom}
{\bf p}=m_0{\bf v}
\end{equation}
is  the  momentum  of  a  particle.
\begin{equation}\label{en}
\displaystyle
{\cal E}=m_0c_0c=m_0{c_0}^2\sqrt{1+\frac{v^2}{{c_0}^2}}, \
T=m_0{c_0}^2\Big(\sqrt{1+\frac{v^2}{{c_0}^2}}-1\Big).
\end{equation}
Let ${\cal E}$ be the relativistic energy,  $T$ be the kinetic  energy
of a particle. The momentum ${\bf p}$, energies ${\cal E}$, $T$ are the
integrals of motion for a free particle. The energy ${\cal E}$ and the
momentum ${\bf p}$ may be united into single 4-momentum (as in SR)
\begin{equation}\label{p}
p^i=m_0c_0u^i=\Big(\frac{m_0c_0}{\sqrt{1-u^2}},\frac{m_0c_0u^{\alpha}}
{\sqrt{1-u^2}}\Big)=\Big(\frac{{\cal E}}{c_0},m_0{\bf v}\Big).
\end{equation}
The components of $p^i$ are related by the ratio:
\begin{equation}\label{4p}
\begin{array}{c}
\displaystyle
p_ip^i=\frac{{\cal   E}^2}{{c_0}^2}-{\bf    p}^2={m_0}^2{c_0}^2;    \\
\displaystyle   {\bf   p}=\frac{{\cal   E}}{c_0c}{\bf   v};         \\
\displaystyle
{\bf p}=\frac{{\cal E}}{c_0c}{\bf c}=\frac{{\cal E}}{c_0}{\bf n}, \ 
{\bf n}=\frac{{\bf c}}{c}, \ if \ m_0=0, \ {\bf v}={\bf c}.
\end{array}
\end{equation}

\begin{sloppypar}\noindent\section{Equations of  Motion  for   Charged
Particle and Electromagnetic Field} \end {sloppypar}\label{5}
\noindent
Let us start from the mechanical [20]  and  field  Lagrange  equations
[21, 22]:
\begin{equation}\label{L}
\displaystyle
\frac{d}{dx^0}\frac{\partial L}{\partial{\bf u}}-
\frac{\partial L}{\partial{\bf x}}=0;                                \ 
\displaystyle
\frac{\partial}{\partial x^k}\frac{\partial{\cal L}}
{\partial(\partial
A_i/\partial   x^k)}-
\frac{\partial{\cal  L}}{\partial  A_i}=0.
\end{equation}
\begin{sloppypar}\noindent
Here $L$ is given by the expression (\ref{2.27});  ${\cal L}$  is  the
density  of  the  Lagrange  function ${\cal L}=-(1/c_0)A_ij^i-(1/16\pi
c_0)  F_{ik}F^{ik}$.  Taking  into   account   the   vector   equality
$\nabla({\bf   a}   \cdot{\bf  b})=({\bf  a}\cdot\nabla){\bf  b}+({\bf
b}\cdot\nabla){\bf  a}+  {\bf  a}{\rm  x}(\nabla{\rm  x}{\bf  b})+{\bf
b}{\rm  x}(\nabla{\rm  x}{\bf  a})$,  the  permutational ratios of the
tensor     of     electromagnetic      field,      the      expression
$\partial(F^{ik}F_{ik})/\partial(\partial  A_i/\partial x^k)=-4F^{ik}$
[20] and relations  (\ref{fg1})  with  $\nabla  c=0$,  we  obtain  the
following equations of motions [17]:
\end{sloppypar}
\begin{equation}\label{2.33}
\frac{d{\bf p}}{dt}=m_0\frac{d{\bf v}}{dt}=\frac{c}{c_0}e{\bf E}+
\frac{e}{c_0}{\bf v}{\rm x}{\bf H};
\end {equation}
\begin{equation}\label{2.34}
\frac{d{\cal E}}{dt}=e{\bf E}\cdot{\bf v}\to m_0\frac{dc}{dt}=
\frac{e}{c_0}{\bf v}\cdot{\bf E}.
\end{equation}

\begin{equation}\label{2.36}
\begin{array}{ll}
\vspace{2mm}
\displaystyle
\nabla{\rm X}{\bf E}+\frac{1}{c}\frac{\partial{\bf H}}{\partial t}=0; &
\displaystyle
\nabla\cdot{\bf E}=4\pi\rho;                                          \\
\displaystyle
\nabla{\rm X}{\bf H}-\frac{1}{c}\frac{\partial{\bf E}}{\partial t}=
\displaystyle
4\pi\rho\frac{{\bf v}}{c}; & \nabla\cdot{\bf H}=0.
\end{array}
\end{equation}
\begin{sloppypar} \noindent
The equations (\ref{2.33})-(\ref{2.36}), considered as the whole, form
the  set  of  nonlinear  electrodynamics  equations which describe the
joint motion of an electrical charge and electromagnetic field. Taking
into account the expression for the velocity of light
\end{sloppypar}
\begin{equation}\label{cv}
\begin{array}{c}
\displaystyle
c(t)=c_0\sqrt{1+\frac{v^2(t)}{{c_0}^2}}=c(0)\Big[1+\frac{e}{m_0c_0c(0)}
\int\limits_{0}^{t}{\bf v}\cdot{\bf E}d\tau\Big]=                   \\
\displaystyle
c(0)\Big[1+\frac{{\cal E}(t)-{\cal E}(0)}{m_0c_0c(0)}\Big]=
c(0)\frac{{\cal E}(t)}{{\cal E}(0)},
\end{array}
\end{equation}
where $c(0)$ is the velocity of light at $t=0$, ${\cal E}=m_0c_0c(0)$,
we may rewrite the set (\ref{2.36}) in the following form
\begin{equation}\label{set}
\begin{array}{ll}
\vspace{1mm}
\displaystyle
m_0\frac{d{\bf v}}{dt}=
\frac{c(0)}{c_0}\Big[1+\frac{ {\cal E}(t)-{\cal E}(0) }{{\cal E}(0)}
\Big]e{\bf E}+\frac{e}{c_0}{\bf v}{\rm x}{\bf H};&{}                \\
\vspace{3mm}
\displaystyle
\frac{d{\cal E}}{dt}=e{\bf v}\cdot{\bf E}\to{\cal E}(t)-{\cal E}(0)=
e\int\limits_{0}^{t}{\bf v}\cdot{\bf E}d\tau;             &  {}     \\
\vspace{1mm}
\displaystyle
\Big[1+\frac{ {\cal E}(t)-{\cal E}(0) }{{\cal E}(0)}\Big]\nabla{\rm X}
{\bf E}+\frac{1}{c(0)}\frac{\partial{\bf H}}{\partial t}=0;          &
\displaystyle
\nabla\cdot{\bf E}=4\pi\rho;                                        \\
\displaystyle
\Big[1+\frac{{\cal E}(t)-{\cal E}(0)}{{\cal E}(0)}\Big]
\nabla{\rm X}{\bf H}-\frac{1}{c(0)}\frac{\partial{\bf E}}{\partial t}=
4\pi\rho\frac{{\bf v}}{c(0)};                                        &
\nabla\cdot{\bf H}=0.
\end{array}
\end{equation}
They coincide  with  the  well-known  SR   equations   [20]   in   the
approximation   $[{\cal  E}(t)-{\cal  E}(0)]/{\cal  E}(0)\ll  1$  when
$c(0)=c_0$.

\begin{sloppypar}\section{Transformational Properties  of  3-Velocity,
Momentum, Energy and Electromagnetic Field}\end{sloppypar}

Let us start from the infinitesimal transformations  (\ref{f2}).  They
induce   the   transformations   of  4-velocity  ${u'}^i={L^i}_ku^k, \
u^k=dx^k/ds$.  As consequence, the formulas of transformations for the
3-velocity take the form:
\begin{equation}\label{2.7}
\begin{array}{lll}
\vspace{1mm}
\displaystyle
\frac{v_x'}{c'}=\frac{\frac{v_x}{c}-\frac{V}{c}}{1-\frac{v_x V}{c^2}};&
\vspace{1mm}
\displaystyle
\frac{v_y'}{c'}=\frac{v_y}{c}\frac{\sqrt{1-\frac{V^2}{c^2}}}
{1-\frac{v_x V}{c^2}};                                                &
\vspace{1mm}
\displaystyle
\frac{v_z'}{c'}=\frac{v_z}{c}\frac{\sqrt{1-\frac{V^2}{c^2}}}
{1-\frac{v_x V}{c^2}};
\end{array}
\end{equation}
\begin{equation}\label{2.7a}
\begin{array}{c}\vspace{2mm}
\displaystyle
\sqrt{1-\frac{v'^2}{c'^2}}=\frac{\sqrt{1-\frac{v^2}{c^2}}\sqrt{1-\frac{V^2}
{c^2}}}{1-\frac{v_x V}{c^2}};   \
\displaystyle
\frac{v'^2}{c'^2}=\frac{\frac{({\bf v}-{\bf V})^2}{c^2}-\frac{({\bf V}{\rm x}
{\bf v})^2}{c^4}}{\biggl(1-\frac{{\bf V}\cdot{\bf v}}{c^2}\biggr)^2}.
\end{array}
\end{equation}
In accordance with these results, the velocity $v$ does not exceed the
velocity of light $c$,  i.e.  $v\le c$ and analogously $v'\le c'$. The
velocity of light $c'$ in the reference frame $K'$ corresponds to  the
velocity of  light  $c$  in the frame $K$ (as in SR).  If $c'=c$,  all
these formulas  go  into  SR  formulas  [20].  Bearing  in  mind  that
$c'=c(1-Vv_x/c^2)/(1-V^2/c^2)^{1/2}$, we find   the   transformational
properties of 3-velocity in the present work:
\begin{equation}\label{2.7b}
\begin{array}{lll}
\vspace{1mm}
\displaystyle
v_x'=\frac{v_x-V}{\sqrt{1-\frac{V^2}{c^2}}};                         &
\vspace{1mm}
\displaystyle
v_y'=v_y;                                                            &
\vspace{1mm}
\displaystyle
v_z'=v_z;
\end{array}
\end{equation}
\begin{equation}\label{2.7c}
\begin{array}{c}
\vspace{2mm}
\displaystyle
c'\sqrt{1-\frac{v'^2}{c'^2}}=c\sqrt{1-\frac{v^2}{c^2}};              \
\displaystyle
v'^2=\frac{({\bf v}-{\bf V})^2-({\bf V}{\rm x}{\bf v})^2}
{1-\frac{V^2}{c^2}}.
\end{array}
\end{equation}
The first  formula  (\ref{2.7c})  also  follows  from  the  expression
$(c'^2-v'^2)dt'^2=(c^2-v^2)dt^2={c_0}^2{dt_0}^2$     for    4-interval
$ds^2$, when $dt_0=dt=dt'$ in accordance with (\ref{lt}).

Analogously, from the  expression  ${p'}^i={L^i}_k  p^k$  one  can  be
obtained  the formulas,  describing transformational properties of the
3-momentum ${\bf p}$ and energy ${\cal E}$:
\begin{equation}\label{2.37}
\begin{array}{c}
\displaystyle
{p'}_x=\frac{p_x-V{\cal E}/c_0c}{\sqrt{1-\frac{V^2}{c^2}}},          \
{p'}_y=p_y,  \  {p'}_z=p_z;                                          \
\displaystyle
{\cal E}'=\frac{{\cal E}-Vp_xc_0/c}{\sqrt{1-\frac{V^2}{c^2}}}.
\end{array}
\end{equation}
If here  $c'=c=c_0$,  we have the SR theory formulas [20].

The transformational  properties  of  the density of electrical charge
and  electromagnetic  field  may  be  obtained  from  the  expressions
${j'}^i={L^i}_kj^k$ and  ${F'}_{ik}={L^l}_i{L^m}_kF_{lm}$  in the case
of the  free  motion  of  electrical  charge   when   $c=const$:
\begin{equation}\label{2.39}
\displaystyle
\rho'=\rho\frac{1-\frac{v_x V}{c^2}}{\sqrt{1-\frac{V^2}{c^2}}}.
\end{equation}
\begin{equation}\label{2.41}
\begin{array}{c}
\vspace{1mm}
\displaystyle
{E_x}'=E_x; \ {E_y}'=\frac{E_y-\frac{VH_z}{c}}{\sqrt{1-\frac{V^2}{c^2}}};          \
\displaystyle
{E_z}'=\frac{E_z+\frac{VH_y}{c}}{\sqrt{1-\frac{V^2}{c^2}}};         \\
\displaystyle
{H_x}'=H_x; \ {H_y}'=\frac{H_y+\frac{VE_z}{c}}{\sqrt{1-\frac{V^2}{c^2}}};          \
\displaystyle
{H_z}'=\frac{H_z-\frac{VE_y}{c}}{\sqrt{1-\frac{V^2}{c^2}}}.
\end{array}
\end{equation}
Formally they are common both for SR theory [20] and for the present work.  
The expression (\ref{2.39})  may   be   also   written   in   the   form
$\rho'/c'=\rho/c$.  Using the method of replacement of variables,  the
transformational properties of electromagnetic field can be applied to
prove the invariance of Maxwell equations (\ref{2.36}) with respect to
space-time transformations (\ref{if3}).

\section {Energy and Faster-than-Light Motion}
Let us     begin     with      the      expression      $v=\sqrt{{\cal
E}^2-{m_0}^2{c_0}^4}/m_0c_0  > c_0$.  It follows from here that in the
framework  of  the  present  work  a  particle  will   move   at   the
faster-than-light  velocity,  if  the  particle  energy  satisfies the
condition:
\begin {equation}\label{g54}
{\cal E}>\sqrt2 {\cal E}_0=\sqrt2 m_0 {c_0}^2.
\end{equation}
(${\cal E}_0=m_0{c_0}^2$).  The energy ${\cal E}=\sqrt2 {\cal E}_0$ is
equal $\sim 723~{\rm keV}$ for the electron (${\cal E}_0\sim  510~{\rm
keV}$)  and  $\sim  1330~{\rm Mev}$ for the proton and neutron (${\cal
E}_0\sim 938~{\rm MeV}$).  From here  we  may  conclude  that  in  the
present work the neutron physics of nuclear reactors may be formulated
in the approximation $v<c_0$ as in SR.  The electrons with the  energy
${\cal  E}>723~{\rm  keV}$  would  be faster-than-light particles (for
example,  the velocity of the $1~{\rm GeV}$ electron  would  be  $\sim
2000~c_0$);  the  particle  physics on accelerators with the energy of
protons  more  than  $1.33~{\rm  GeV}$  would  be   the   physics   of
faster-than-light motions  if  the present theory were realized in the
field of  validity  of  SR.  The  examples  of  using  the  space-time
transformations (\ref{if3})   for  the  interpretation  of  Michelson,
Fizeau and some other experiments,  as well as for the  interpretation
of aberration  of  light  and  of  Doppler  effect,  decay of unstable
particles and creation  new  particles,  faster-than-light  motion  of
nuclear reaction products are given in [17].

\section{Conclusion}
The $L_6{\rm X}\triangle_1$ invariant  theory  has  been  constructed,
where   $L_6$  is  the  Lorentz  group,  $\triangle_1$  is  the  scale
transformation group of the velocity of light $c'=\gamma c$. The field
of  application  of  the  theory  is  yet unknown in the present time.
Nevertheless in according with  the  Blokhintsev  papers  [1]  we  may
assume  that  the proposed theory will prove to be useful in the field
of quantum physics of dimensional particles,  where  the  property  of
elementary  nature  should not contradict to the existence of internal
structure of the particle.  Indeed, the elementary particles should be
points  in  the $L_6$ invariant theory (SR) because of a finiteness of
the speed of light $c_0$.  In the  $L_6{\rm  X}\triangle_1$  invariant
theory this requirement is not necessary because of the absence of the
limit to the velocity of light $c$.  The postulation $c'=c$  leads  to
SR. \bigskip

\section{Acknowledgement}

The author   is   deeply   grateful   to   Prof.~A.E.~Chubykalo    and
Prof.~V.V.~Dvoeglazov   of   Zacatecas   University  for  the  helpful
discussion and to Prof.~A.E.~Chubykalo for the invitation  to  publish
in the Present Issue.

\nonumsection{References}


\begin{thebibliography}{000}

\bibitem{1} D.~I.~Blokhintsev,      "Note      on     the     Possible
Relativistic-Invariant Generalization of the Concept of Field",  JETP,
{\bf 16},  480-482 (1946);  "On a non-Hamiltonian Method in the Theory
of Elementary Particles", JETP, {\bf 17}, 266-271 (1947).

\bibitem{2} D.~Blokhintsev,   "On   the   Propagation  of  Signals  in
Nonlinear Theory of Field",  Doklady Akad. Nauk, {\bf LXXXII}, 553-556
(1952).

\bibitem{3} D.~A.~Kirzhnits,   "To   the   Question  on  Meson-Nucleon
Interactions", JETP, {\bf 27}, 6-18 (1954).

\bibitem{4} Ya.~P.~Terletsky,  "Principle of Causality and the Second
Law of Thermodynamics", Doklady Akad. Nauk, {\bf 133}, 329-332 (1960).

\bibitem{5} G.~Feinberg,  "On  the  Possibility  of Faster than Light
Particles", {\it Einstein Sbornik  1973}  (Moscow,  Nauka,  1974)  pp.
134-177; From: Phys. Rev., {\bf 159}, 1089 (1967).

\bibitem{6} O.~Bilaniuk,  E.~Sudarshan,  "Particles Beyond  the  Light
Barrier", see Ref.  5,  pp. 112-133; From: Physics Today, {\bf 22}, 43
(1969).

\bibitem{7} E.~Recami,  {\it Relativity Theory and its Generalization.
Astrophysics,  Quanta and Relativity Theory} (Moscow, Izdatelstvo Mir,
1982) pp. 53-128.

\bibitem{8} R.~Mignani,  "Quaternionic  Form  of  Superluminal Lorentz
Transformations", Lett. Nuovo Cimento, {\bf 13}, 134-138 (1975).

\bibitem{9} D.~A.~Kirzhnits,  V.~N.~Sazonov,  "Superluminal Motions in
Special Relativity Theory", see Ref. 5, pp. 84-111.

\bibitem{10} H.~C.~Corben,  {\it  Tachyons,  Monopolies,  and  Related
Topics} (Ed. by Recami~E. North-Holland Publ. Co., Amsterdam, 1978) p.
31, see Ref. 7, p. 124.

\bibitem{11} C.~E.~Patty,  "Electromagnetic  Behavior  in Superluminal
Interactions: the Classical Electromagnetic Problem", Nuovo Cim., {\bf
70B}, 65-79 (1982).

\bibitem{12} Erasmo~Recami,     Flavio~Fontana,    Roberto~Caravaglia,
"Special Relativity and Superluminal Motions:  a  Discussion  of  Some
Recently  Experiments",  Int.  J.  of  Modern  Physics  A.,  {\bf 15},
2793-2812 (2000).

\bibitem{13} W.~Pauli,  {\it Relativit\"atstheorie, Encyklop\"adie der
Mathematischen Wissenschaften,  Band~{\bf 2}, Heft~IV, Art.~19, 1922},
ss.~539--775;    in    Russian:    {\it    Theory    of    Relativity}
(Moscow-Leningrad, Gostexizdat, 1947) pp.~24; 274; 282.

\bibitem{14} A.~A.~Logunov,   {\it   Lections   on   Fundamentals   of
Relativity Theory} (M.~V.~Lomonosov  Moscow  State  University,  1982)
pp.~20-40, 63, 64-85 (in Russian).

\bibitem{15} Sheldon~L.~Glashow,  "How  Cosmic-Ray Physicists Can Test
Special Relativity",  Nucl. Phys., B (Proc. Suppl.), {\bf 70}, 180-184
(1999).

\bibitem{16} P.~M.~Rapier,  "An  Extension  of Newtonian Relativity to
Include Electromagnetic Phenomena",  Proc.  IRE,  {\bf 49},  1691-1692
(1961); "A Proposed Test for Existence of a Lorentz-Invariant Aether",
IRE, {\bf 50}, 229-230 (1962).

\bibitem{17} G.~A.~Kotel'nikov,  "On the Invariance of the Velocity of
Light in Special Relativity Theory",  Vestnik of M.V.~Lomonosov Moscow
State  University,  Physics  \&  Astronomy,  No.~4,  371-373   (1970);
"Lorentz-Invariant  Theory  Permitting  Superluminal  Motion",  J.  of
Russian Laser Research, {\bf 22}, 455-474 (2001).

\bibitem{18} J.~P.~Hsu, "New Four-Dimensional Symmetry", Found. Phys.,
{\bf  6},  317-339  (1976);  J.~P.~Hsu  and Leonardo~Hsu,  "A Physical
Theory Based on the First Postulate of  Relativity",  Phys.  Lett.  A,
{\bf 196}, 1-6 (1994).

\bibitem{19} Andrew~E.~Chubykalo,  Roman~Smirnov-Rueda,  "Action  at a
Distance as a Full-Value Solution of Maxwell Equations:  the Basis and
Application of the Separated-Potentials Method",  Phys.  Rev.  E, {\bf
53}, 5373-5381 (1996).

\bibitem{20} L.~D.~Landau  and  E.~M.~Lifshitz.  {\it  The  Theory  of
Field} (Moscow, Fizmatgiz, 1973) pp. 66-71, 93-104, 110 (in Russian).

\bibitem{21} D.~Ivanenko  and  A.~Sokolov.  {\it  Classical  Theory of
Field} (Moscow-Leningrad, Gostexizdat, 1951) pp.~135-145 (in Russian).

\bibitem{22} N.~N.~Bogoliubov and D.~V.~Shirkov.  {\it Introduction to
the  Theory  of  Quantized  Fields} (Moscow,  Nauka,  1973) p.  76 (in
Russian).


\end{thebibliography}
\end{document}